\documentclass[pre,showpacs,superscriptaddress,floatfix]{revtex4}
\usepackage{epsfig}
\begin{document}
\title{Crossover from Reptation to Rouse dynamics in the Cage  Model}
\author{A. Drzewi\'nski}
\affiliation{Czestochowa University of Technology, 
Institute of Mathematics and Computer Science,
ul.Dabrowskiego 73, 42-200 Czestochowa, Poland}
\author{J.M.J. van Leeuwen}
\affiliation{Instituut-Lorentz, Leiden University, P.O. Box 9506,
2300 RA Leiden, The Netherlands}

\date{\today}

\begin{abstract}
The two-dimensional cage model for polymer motion is discussed with an emphasis on the
effect of sideways motions, which cross the barriers imposed by the lattice. 
Using the Density Matrix Method as a solver of  the Master Equation, the renewal time and the 
diffusion coefficient are calculated as a function of the strength of the barrier crossings. 
A strong crossover influence of the barrier crossings is found and it is 
analyzed in terms of effective exponents for a given chain length. The
crossover scaling functions and the crossover scaling exponents are calculated.
\end{abstract} 

\pacs{05.10.-a, 83.10.Kn, 61.25.Hq}

\maketitle

\section{Introduction}

The motion of  a single polymer dissolved in a gel has been described in terms of reptation
\cite{deGennes}.  Typical for reptation is that the polymer chain moves inside a tube, 
which can only be refreshed by growth and shrinkage at the ends.
In order to make this motion suitable for analysis, lattice models have been designed of which
the cage model, proposed by Edwards and Evans in 1981 \cite{Edwards}, is the oldest. 
It was mainly considered as a simple model for reptation \cite{deGennes,Viovy}, 
which is indeed the main mode of motion for polymers dissolved in a gel. The embedding 
lattice plays the role of the gel imposing barriers for the motion. The original introduction 
allowed for two types of motion: ``reptations'', which are motions along the confining tube 
and sideways motions or ``barrier crossings'', in which the chain overcomes a barrier and 
thus changes the tube configuration. Sofar most attention has been paid to the reptations 
only. In this paper we concentrate on the interplay, which is quite delicate, as we will show.
E.g. naively one might think that the diffusion coefficient is the linear sum of the  
contributions of the two mechanisms, but that is not at all the case, as has been noted by 
Klein Wolterink and Barkema \cite{Barkema2} in a similar context. This makes it difficult to 
analyze experiments, in which never the simultaneous presence of the
two types of motion can be excluded. In the cage model the interplay of the two modes can 
be fully analyzed.

The model has extensively been studied by Monte Carlo simulations, which seems the 
only way to deal with the Master Equation for the stochastic motion 
\cite{Deutsch,Kremer2,Reiter,Barkema,Heukelum}. 
Since the relevant  Master Equation does not obey detailed balance, no systematic solution 
method exists. The issues, to which we presently address ourselves, were the subject of
a lively debat in the early nineties. The various simulations showed certain tendencies,
but the limitation to fairly short chains and the intrinsic statistical noise, prevented in our
opinion definite settling the role of Rouse dynamics vs  reptation.

In this paper we use an alternative method, based on the analogy between the Master 
Equation and the Schrodinger equation, by which the temporal evolution of the probability 
distribution of the chain configurations corresponds to the evolution of the wave function. 
Of course the wave function may be complex, while the probability distribution 
is real and positive. Also the Master operator,
viewed as a hamiltonian, is non-hermitian, which implies decay towards the stationary state
in contrast to the oscillatory temporal behavior of the eigenfunctions of quantum problems. 
Inspite of these differences one can benefit from the analogy, the more so because the
Master operator corresponds to the hamiltonian of a one-dimensional spin chain, for which
the very efficient Density Matrix Method (DMRG) has been designed by White \cite{White}.
The cage model remains a one-dimensional quantum problem, irrespective the lattice in which
it is embedded, because the chain itself is a linear structure. 

We focus on two dynamic properties: the renewal time $\tau$ and the diffusion coefficient 
$D$ and determine them directly from the Master Operator. Both properties refer to 
asymptotically long times (the stationary state) and thus our calculations are complementary 
to the Monte Carlo simulations which probe the short and intermediate time behavior 
\cite{Schafer}.
The renewal time is the time needed for the chain to assume a new configuration,
which has no memory of the original one. It is found from the gap in the spectrum of the
Master Operator. The Master Equation always has a trivial eigenvalue 0, corresponding
to the stationary state. Any other initial state ultimately decays towards the stationary
state and the slowest relaxation time (the inverse of the gap) is the renewal time. Its
calculation is difficult because for long chains, the gap gets very small and the excited
states are hard to disentangle from the stationary state. In fact the gap decays with
as a negative power $z$ of the length $N$ of the chain, such that $\tau \sim N^z$. 
The zero field diffusion coefficient $D$ is related to the drift velocity in a weak driving field. 
It decays as a power $N^{-x}$. 

We will confine ourselves to one- and two-dimensional embeddings.
The degree of difficulty of the solution is related to the dimension 
of the embedding lattice, simply because the higher the embedding dimension, the higher is
the spin in the corresponding spin chain and the more states are required in the 
DMRG approximation. 
In this paper we will show that the one-dimensional version, 
which is admittedly unrealistic, allows an analytical solution. The two-dimensional 
embedding lattice presents the problem already in its full complexity, 
while of course the three-dimensional case is of the most experimental relevance. 
Actually the universal properties are believed to be the same for embedding lattices from
$d=2$ and higher. The practical limitation to two-dimensional embedding lattices derives 
from the fact that our computations are already at the limits of the present day 
possibilities. 

\begin{figure}[h]
\begin{center}
    \epsfxsize=12cm
    \epsffile{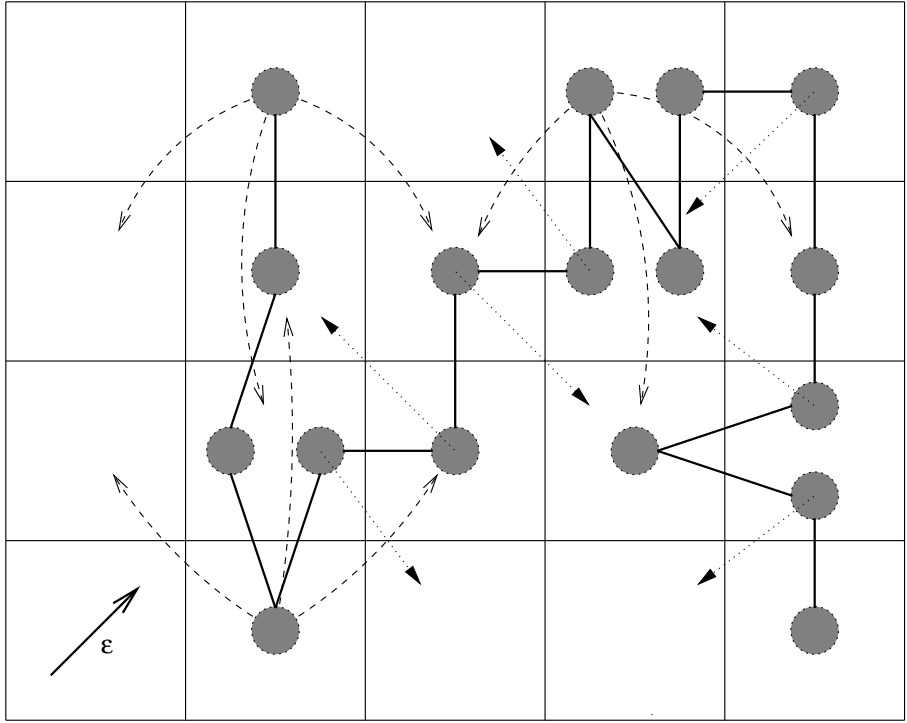}
    \caption{A picture of the polymer chain, consisting of reptons (grey dots) and some 
       	examples of allowed repton motions. The dashed arrows denote reptations 
	(hernia moves), whereas the dotted ones represent barrier crossings. The solid
	 arrow presents the driving field direction.}  \label{chain}
\end{center}
\end{figure}

The DMRG approach yields very accurate results in the domain where it converges. This
enables us to use finite-size scaling analysis for the determination of the exponents and the
crossover scaling functions. The case without crossing barriers 
(the reptation dynamics) has powers different from the case with
these crossings (Rouse dynamics). From the viewpoint of dynamic scaling, the exponents
are exotic and nice illustrations of how crossover takes places from one type of behavior
to an other. A crossover scaling representation for $\tau$ and $D$ strongly elucidates their
behavior.

\section{The model}

The model is a chain of $N+1$ reptons, connected by $N$ links, $(y_1, \cdots, y_N)$, 
to neighboring cells of a hypercubic lattice. A picture of a chain is given in Fig.~\ref{chain}. 
The mobile units are the reptons. The $y_i$ can take $2d$ values 
and because we have a two-dimensional embedding lattice, we 
can denote them by the directions $N$(orth), $E$(ast), $S$(outh) and $W$(est). 
Two consecutive reptons are always in adjacent cells, but the chain may backtrack and 
cells can be multiply occupied, (see Fig.~\ref{chain}).
The $y_i $ variables characterize the chain, since
the absolute position is irrelevant for the properties that we consider. 
The statistics of the model is governed by the Master Equation for the probability 
distribution $P ({\bf Y},t)$, where ${\bf Y}$ stands for the complete 
configuration $( y_1, \cdots, y_N)$. It has the general form
\begin{equation} \label{a1}
{\partial P({\bf Y},t) \over \partial t} =  \sum_{\bf Y'}\left[ W ({\bf Y} | {\bf Y}')
P ({\bf Y}',t) - W ({\bf Y}' | {\bf Y}) P({\bf Y},t)\right]
\equiv \sum_{\bf Y'} M({\bf Y},{\bf Y}') P({\bf Y}',t).
\end{equation}
The $W$'s are the transitions rates of the possible motions that we have indicated in
Fig.~\ref{chain}.
The matrix $M$ combines the gain terms (in the off-diagonal elements) and the loss
terms (on the diagonal). $M$ is a sum of matrices, for each repton one
\begin{equation} \label{a2}
M({\bf Y},{\bf Y}') = \sum^N_{i=0} M_i({\bf Y},{\bf Y}'),
\end{equation}
where the sum runs over the reptons starting with the tail repton $i=0$ to the head
repton $i=N$. The tail repton matrix is diagonal in all the link variables except the first
\begin{equation} \label{a3}
M_0 ({\bf Y},{\bf Y}') = m_0 ( y_1| y'_1) 
\prod^N_{i=2} \delta_{ y_i,y'_i}.
\end{equation} 
The tail repton produces exclusively reptations. Each move of the tail repton can be seen 
as a combination of a withdrawal towards the cell of the next repton and from thereon 
a move to a new cell. The matrix $m_0$ is explicitly given by the scheme
\begin{center}
\begin{tabular}{|c|c|c|c|c|}
\hline 
  & & & & \\*[-2mm]
$ y \setminus  y'$   & $N$  & $ E$  &  $S$  & $ W$  \\*[2mm]
\hline
  & & & & \\*[-2mm]
$N$ & $-1 -2B^2$ & 1 & $B^{-2}$ & $B^{-2}$ \\*[2mm]
\hline
 & & & & \\*[-2mm]
$E$ & 1 &$-1 -2B^2$   &$B^{-2}$   &$B^{-2}$ \\*[2mm]
\hline
 & & & & \\*[-2mm]
$S$ & $B^2$  & $B^2$ & $-1 -2B^{-2}$  & 1 \\*[2mm]
\hline
 & & & & \\*[-2mm]
$W$ & $B^2$ &$B^2$  & 1 &$-1 -2B^{-2}$   \\*[2mm]
\hline
\end{tabular}
\end{center}

The parameter $B=\exp (\epsilon/2)$ is a bias, which accounts for the influence of a
driving field, which can be an electric field when the reptons are charged. The value of 
(the small) $\epsilon$ is a dimensionless measure for the 
strength of this driving field. The driving field is along the body diagonal, here in the 
North-East direction. So if the link $N$ moves to the direction $W$, the tail repton moves 
\begin{figure}[h]
\begin{center}
    \epsfxsize=12cm
    \epsffile{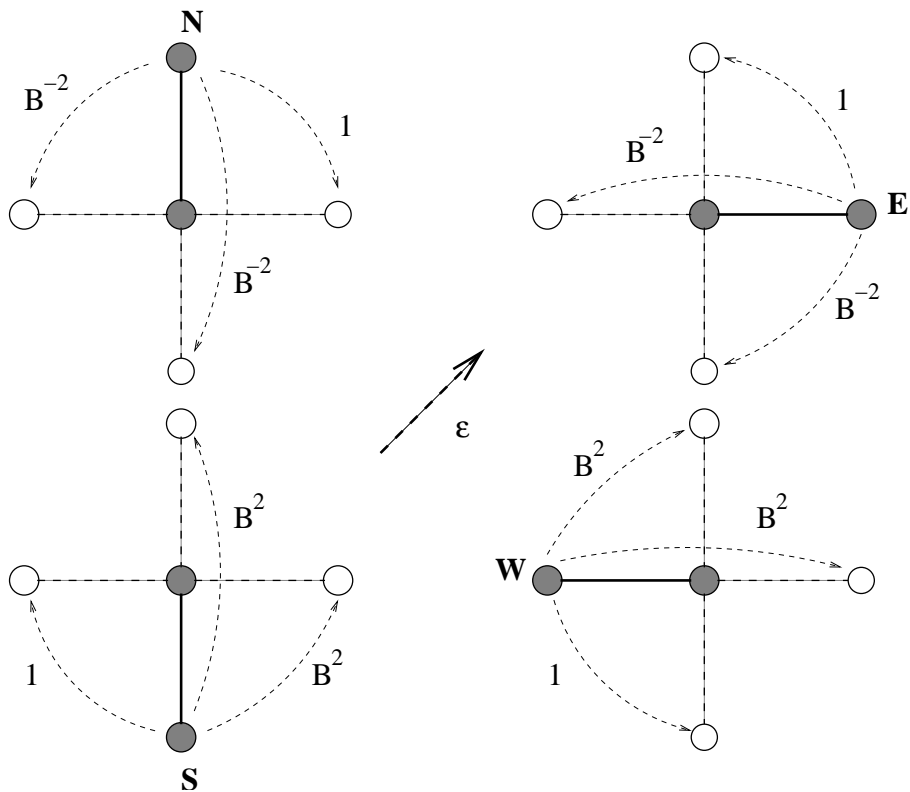}
    \caption{The allowed motions of the head repton with corresponding transition ratios. 
    The dashed arrow presents the driving field.}  \label{head}
\end{center}
\end{figure}
two units in the direction of the field. The reverse process gets a bias $B^{-2}$. The head
repton transition probabilities are given by a similar matrix with $B^2$ replaced by $B^{-2}$.
They are depicted in Fig.~\ref{head}.
One could give all the transitions an overall factor, but this would only influence the
overall time rate. So we keep the unbiased transitions equal to 1.

\begin{figure}[h]
\begin{center}
    \epsfxsize=12cm
    \epsffile{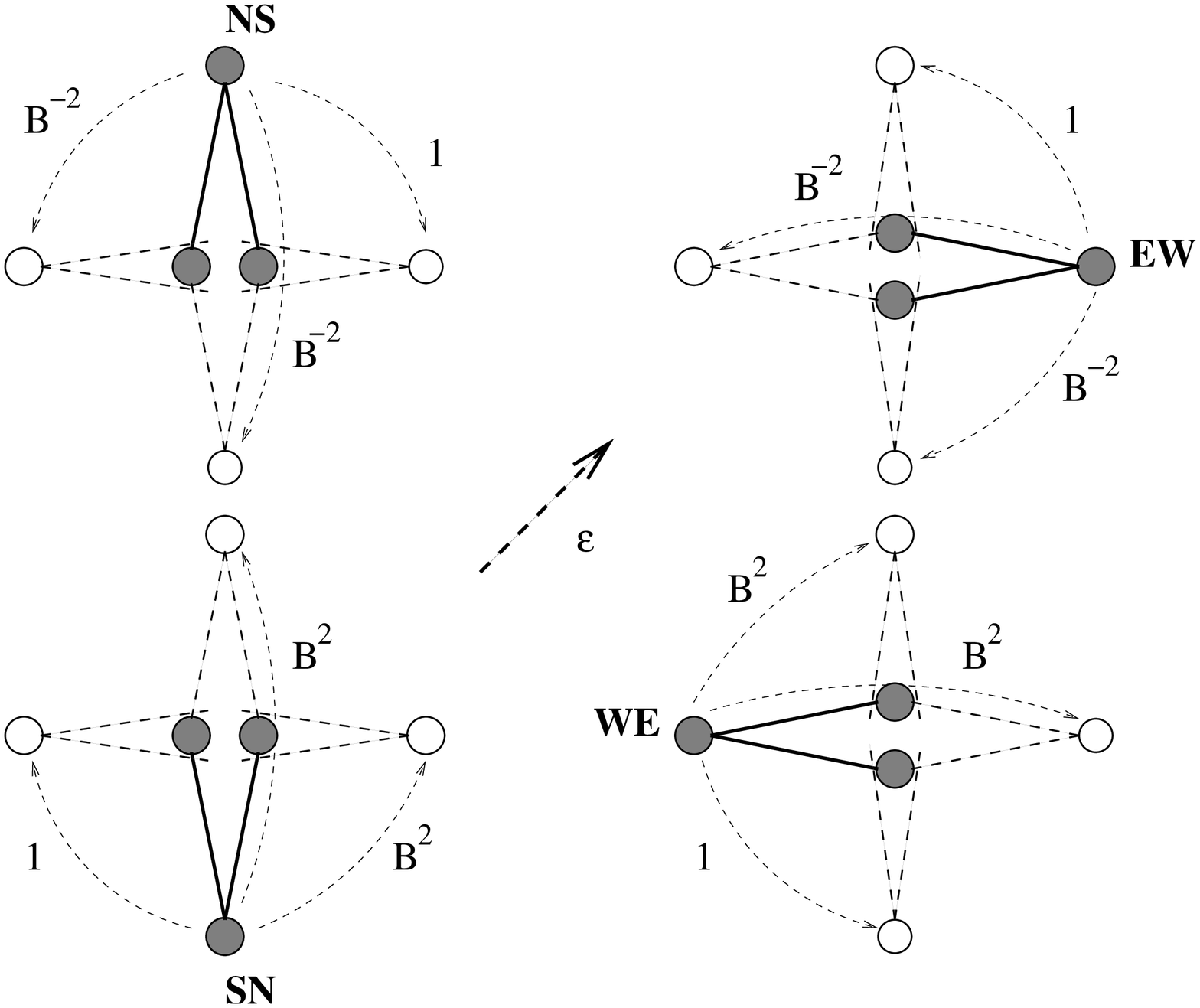}
    \caption{The allowed motions of the hernia with corresponding transition ratios. 
    The dashed arrow presents the driving field.}  \label{hernia}
\end{center}
\end{figure}

\begin{figure}[h]
\begin{center}
    \epsfxsize=12cm
    \epsffile{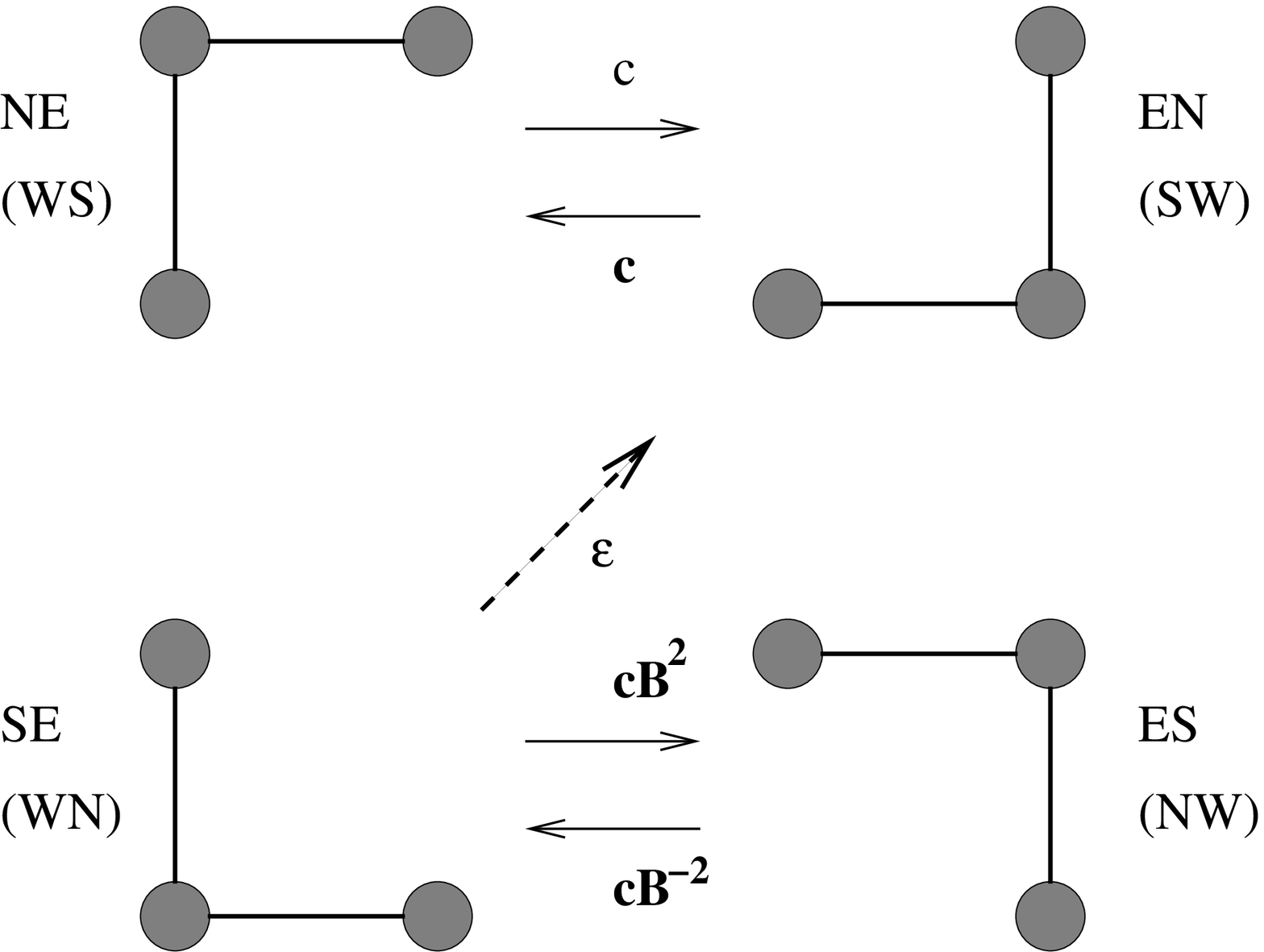}
    \caption{The solid arrows present allowed motions for the barrier crossing with 
      corresponding transition ratios. The dashed arrow presents the driving 
      field.}\label{kink}
\end{center}
\end{figure}

The internal repton $i$  changes two consecutive links $y_i,y_{i+1}$. 
The matrix $M_i$ is diagonal in all the other link variables. The transition matrix 
contains two types of contributions:
\begin{itemize}
\item {\it Reptations.} These are the cases where $y'_i$ and $y'_{i+1}$ are opposite, e.g. 
$N$ and $S$ (sometimes called a hernia in the chain). Then repton $i$ can retract to the 
cell to which it is doubly connected and then recreate a new hernia, a new pair of opposite 
links, e.g. $E$ and $W$ (see Fig.~\ref{hernia}). Note that the sequence $EW$ differs from 
$WE$. Moves towards $EW$ and $WE$  get different biases.
\item {\it Barrier crossings.} A typical case is the sequence $NE$. It may flip to the 
sequence $EN$. To reach that new position it necessarily has to cross the lattice point 
enclosed by the two sequences. We give these transitions a (small) factor $c$ together 
with the biases which measure the distance that repton $i$ travels in the direction of 
the field (see Fig.~\ref{kink}).
\end{itemize}

It is worth noticing that for the cage model all transition ratios are proportional to
the bias $B^2$ (or $B^{-2}$) \cite{footnote}. 

Generally, conservation of probability follows from the fact that the sum 
over the columns of the matrix $M$ vanishes. So $M$ has a zero eigenvalue and the 
eigenfunction corresponding to this eigenvalue is the stationary state of the system,
to which every other initial state ultimately decays. The matrix is non-symmetric,
due to the bias, which gives different rates to a process and its inverse. Thus one has 
to distinguish between left and right eigenfunctions. The left eigenfunction, belonging
to the zero eigenvalue, is trivial (all components equal), the right eigenfunction is the
stationary state distribution.

The renewal time is usually defined for bias $B=1$ (no driving field). For the diffusion
coefficient we introduce a small driving field which induces an overall drift 
$v_d \sim \epsilon$. The proportionality coefficient gives the diffusion coefficient $D$ 
according to the Einstein relation
\begin{equation} \label{a4}
D = {1 \over N} \,\left( {\partial v_d \over \partial \epsilon }\right)_{\epsilon=0}. 
\end{equation} 

$M({\bf Y},{\bf Y}')$ is the matrix representation of the Master Operator ${\cal M}$, 
acting as a hamiltonian. As $M$ is non-symmetric the hamiltonian ${\cal M}$ is non-hermitian. 
One may view the states $y_i$ of the links as the states of a discrete plane rotator. 
However, translating the action of ${\cal M}$ in terms of rotator operators, does
not lead to a more transparent expression (except for $d=1$, which maps on a spin $1/2$
chain, see Section \ref{deq1}).

The model has only a few parameters: the length of the chain $N$, the strength of the
driving field $\epsilon$ and the relative strength $c$ of the barrier crossings with respect
to the reptating transitions. Experimentally the most interesting combination is the
case where $\epsilon \rightarrow 0$ and $N \rightarrow \infty$. This is a delicate
limiting process since the product $\epsilon N$ may stay finite and influences the nature
of the stationary state strongly. The properties that we consider: renewal time $\tau$ and
diffusion coefficient $D$, refer to the case where this product remains infinitesimal. 
Thus effectively we have only $N$ and $c$ as parameters. 
We will see that also in this pair interesting scaling combinations occur.

\section{Symmetries of the Master Operator}

For our analysis optimal use of the symmetries of the Master Operator is vital. We have
chosen the driving field in the North-East direction in order to make the directions
$N$ and $E$ as well as $S$ and $W$ equivalent. We use two symmetry operations of the
lattice: 
\begin{itemize}
\item Reflection with respect to the field axis. This turns the direction $N$ into $E$ and 
vice versa and similarly it interchanges $S$ and $W$. We refer to this operation as 
${\cal S}_\perp$.
\item Reflection parallel to the field axis, which interchanges the directions $N$ and $W$
as well as $E$ and $S$. It is denoted as ${\cal S}_\parallel$.
\end{itemize}
The hamiltonian is invariant under the operation ${\cal S}_\perp$ but not under 
${\cal S}_\parallel$. It would, if it were accompanied by a field inversion.
We can analyze the consequences of the symmetry by considering the following 
transformation 
\begin{equation} \label{b1}
\left( \begin{array}{c} 0 \\ 1 \\ 2 \\ 3 \end{array} \right) =  {\cal R} 
\left( \begin{array}{c} N \\ E \\ S \\ W \end{array} \right) = {1 \over 2} \left(
\begin{array}{c} N+E+S+W \\ N-E+S-W \\ N+E-S-W \\ N-E-S+W \end{array} \right)
\end{equation}
${\cal R}$ is an orthogonal transformation of the states $N,E,S$ and $W$ of the links.  
The new states have a definite symmetry character under the two operations.
The state 0 is even under ${\cal S}_\perp$ and ${\cal S}_\parallel$, the state 1 is 
odd and even, 2 is even and odd and 3 is odd for both. Let us illustrate 
the effect of this rotation by applying it to the matrix $m_0$, yielding
\begin{equation} \label{b2}
{\cal R} m_0 {\cal R}^{-1} = \left(  
\begin{array}{c} 0 \\ 0 \\ m_{20}\\ 0 \end{array}
\begin{array}{c} 0 \\ m_{11} \\ 0 \\ m_{31} \end{array}
\begin{array}{c} 0 \\ 0 \\ m_{22}\\ 0 \end{array}
\begin{array}{c} 0 \\ m_{13} \\ 0 \\ m_{33}  \end{array}
\right).
\end{equation}
The entries are now labeled by the states 0,1,2 and 3 and are given by
\begin{equation} \label{b3}
\left\{ \begin{array}{rclclcl }
m_{11} & = & - 2 -(B^2+B^{-2}) & = & m_{33}, \quad m_{22} & = & -2(B^2 + B^{-2}), \\*[2mm] 
m_{20} & = & -2(B^2 - B^{-2}) &   =  & 2 m_{31} & = & 2 m_{13}. \\*[2mm]
\end{array} \right.
\end{equation} 
One sees that the matrix is block diagonal with a $2 \times 2$ matrix in the $0-2$ channel 
and one in $1-3$ channel. This results from the invariance of the hamiltonian with
respect to ${\cal S}_\perp$, since the states 0 and 2 are even and the 
states 1 and 3 are odd under this symmetry. So states of different parity 
under ${\cal S}_\perp$ are not mixed by the hamiltonian. 
One also observes that the matrix becomes diagonal for $B=1$. The off-diagonal 
elements concern transitions where the symmetry under field inversion is changed. 

The symmetry character of the links in the new states can be carried over to larger 
segments of the chain, simply by multiplying the parities of the constituing links. The
states 0,1,2 and 3 not only function as states of the links, but also as indices for the
4 different symmetry classes. What has been shown in detail for the tail repton, holds
also for the total hamiltonian. It is invariant under ${\cal S}_\perp$ and for $B=1$ also
under ${\cal S}_\parallel$.  

For the diffusion coefficient we expand the Master Equation in powers of $\epsilon$. 
\begin{equation} \label{b4}
{\cal M} = {\cal M}_0 + \epsilon {\cal M}_1 + \cdots , \quad \quad \quad 
P ({\bf Y}) = P_0 ({\bf Y}) + \epsilon P_1 ({\bf Y}) + \cdots 
\end{equation} 
and  obtain the equations
\begin{equation} \label{b5}
{\cal M}_0 P_0 = 0 , \quad \quad \quad 
{\cal M}_0 P_1 = - {\cal M}_1 P_0.
\end{equation} 
The first equation is trivially fulfilled by a constant $P_0 ({\bf Y})$, since the matrix 
$M_0$ is symmetric and the right eigenvector becomes equal to the trivial left eigenvector.
Note that $P_0 ({\bf Y})$ is the direct product of state 0 for all the links.
When we perform the rotation (\ref{b1}) on all links, the vector $P_0$ changes from a
constant in all entries into the vector with a 1 in the first entry and a 0 in all others.
 
The second equation is a set of homogeneous linear equations for the components of
$P_1 ({\bf Y})$. It is soluble, since the right hand side of the equation is perpendicular 
to the left eigenvalue (which remains true for all orders in $\epsilon$). So we can make the
solution definite by requiring that it is also orthogonal to the trivial left eigenvector.
$P_1 ({\bf Y})$ yields the lowest order drift velocity $v_d$.   

From the viewpoint of symmetries, the operator ${\cal M}_0$ is invariant with respect
to both ${\cal S}_\perp$ and ${\cal S}_\parallel$. So it does not affect the symmetry 
character of $P_1$, which therefore inherits its symmetry from the right hand side of 
(\ref{b5}). The latter derives its symmetry from ${\cal M}_1$ since $ P_0$ is fully 
symmetric. From the example (\ref{b2}) we can see what the tail-repton hamiltonian does to
the first link: it turns it from state 0 (the building block of $P_0$) to state 2. 
Detailed calculations show that this also holds for the other components of ${\cal M}_1$. 
By ${\cal M}_1$ the vector $P_0$ turns from sector 0 to sector 2. This
is not surprising since closer inspection of the right hand side of (\ref{b5}) 
shows that it is the microscopic expression for the drift velocity.
This reverses sign under ${\cal S}_\parallel$ but stays invariant under ${\cal S}_\perp$, 
which is indeed the symmetry character of sector 2.

\begin{figure}[h]
\begin{center}
    \epsfxsize=12cm
    \epsffile{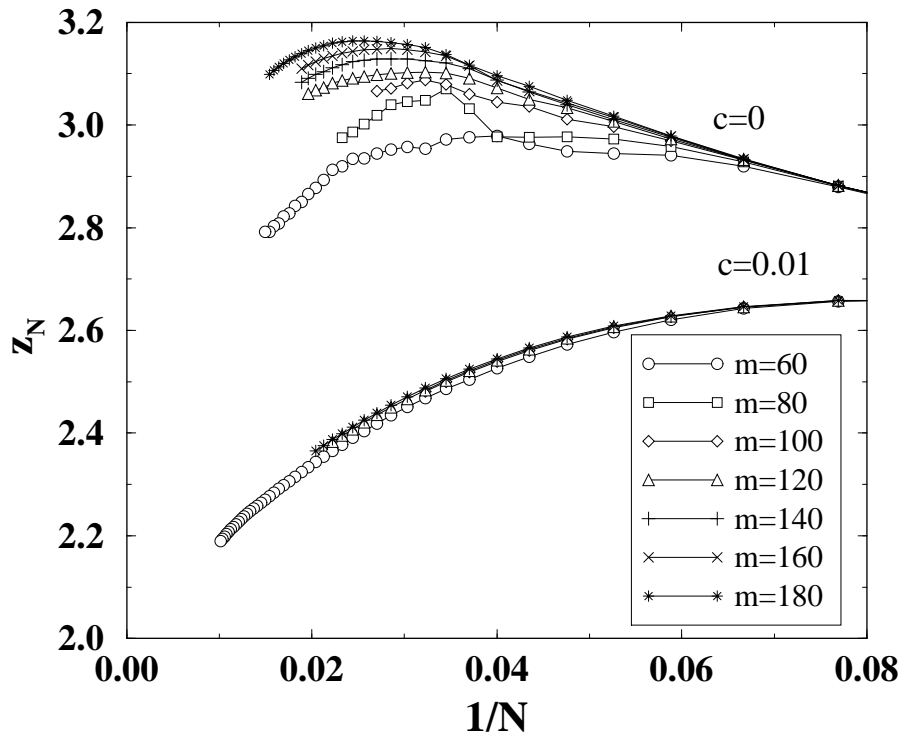}
    \caption{Comparison of m-dependence for the gap exponent at zero and 
    non-zero $c$.}  \label{figCG06}
\end{center}
\end{figure}

\section{The one-dimensional Model} \label{deq1}

As a curiosity we mention the cage model for a one-dimensional embedding lattice.
Then the $y_i$ can take only two values, which one would naturally take as 1 and -1. 
1 is a step forward and -1 a step backward along the embedding line. On the line there 
is no possibility of barrier crossing and we have therefore only reptations. An internal
move takes place when a pair of links (1,-1) turns into (-1,1) or vice versa.
In addition the tail and head link can change from 1 to -1 or from -1 to 1. 
Such a model belongs to the class of models which can be solved by the matrix product 
representation designed by Derrida et al. \cite{Derrida}. Usually the model is discussed
in terms of the variables 1 (a particle) and 0 (a vacancy) and one visualizes the 
dynamics as a form of traffic. The particles cannot overtake each other and have to
wait till they can exchange with a vacancy. The mutual exclusion of particles corresponds 
in the cage model to the fact that a repton cannot move when it is surrounded by two links 
of the same value. At the ends of the chain particles enter and leave with certain rates, 
which in the cage model means that the tail (and head) link can change into their 
opposite direction. Comparing the rules by which 1 and -1 can interchange
in the cage model and the rates at which 1's and -1's are created at the head and tail
of the chain, one finds the rules for the equivalent traffic model.

\begin{figure}[h]
\begin{center}
    \epsfxsize=12cm
    \epsffile{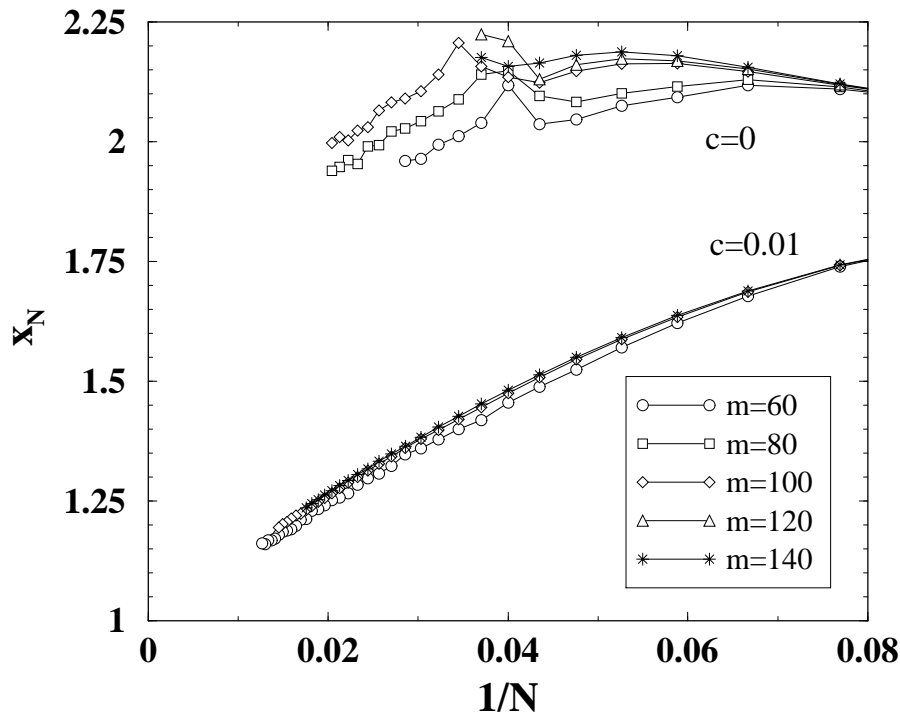}
    \caption{The comparison of m-dependence for the diffusion exponent at $c=0$ and 
    $c=0.01$.}  \label{figCG07}
\end{center}
\end{figure}

The class of such traffic models has been solved by Sasamoto \cite{Sasamoto} and 
independently by  Blythe et al. \cite{Blythe}. 
In \cite{Drzewinski} this model has been related to the
necklace model for reptation, where the results of the traffic model are formulated in terms
of the language of polymer motion and expressions are derived for the drift velocity 
and the diffusion coefficient. It is interesting that one has a solution for the
whole range of values of the bias $B$. E.g. the value of the drift velocity becomes 
independent of the length $N$ and reads for large $N$
\begin{equation} \label{c1}
v_d \simeq {1 \over 4 } \, (B - B^{-1}). 
\end{equation} 
Thus indeed for small $\epsilon$ the drift velocity becomes proportional to $\epsilon$. 
The diffusion coefficient follows using (\ref{a4}) with an asymtotic decay as  
$D \sim N^{-1}$ and the diffusion exponent equals $x=1$.

\begin{figure}[ht]
\begin{center}
    \epsfxsize=12cm
    \epsffile{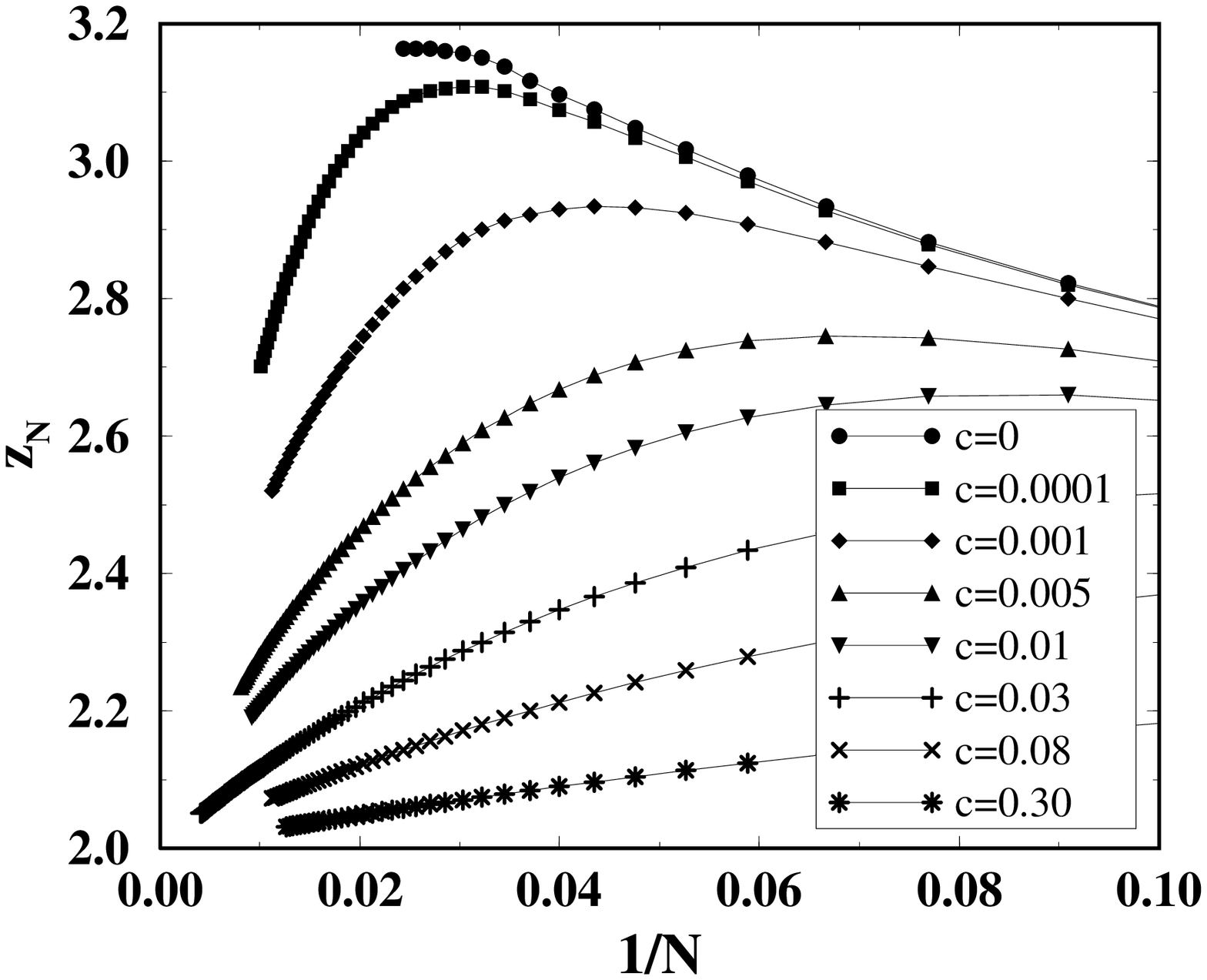}
    \caption{The renewal time exponent as a function of the length of the chain for various 
	values of the barrier crossing rate $c$. Note that we have shortened the curve 
	for $c=0$ with respect to the best curve ($m=180$) in Fig.~\ref{figCG06}, 
	because for longer chains we have
	significant changes with increasing $m$.}  \label{figCG01}
\end{center}
\end{figure}

Not only the gap can be calculated explicitly, but also the whole spectrum of the 
zero field hamiltonian, since it becomes equivalent with the Heisenberg ferromagnetic 
spin chain. The gap $\Delta$ reads
\begin{equation} \label{c2}
\Delta = - 2 (1 - \cos (\pi/N)).
\end{equation} 
Note that the gap vanishes as $\Delta \sim N^{-2}$ for long chains, yielding $z=2$.

Although all the moves of the $d=1$ cage model are reptations, the exponents for the
gap and the diffusion coefficient are not typical for reptation. The reason is that the 
one-dimensional model has no obstacles which slow down the motion by an 
order of magnitude in the chain length $N$, as is characteristic for reptation.

\section{Effective exponents and convergence}

The DMRG expands the solution of the Master Equation in a basis of size $m$. The accuracy
of the method can be tested by an internal parameter: the truncation error \cite{White}, but more 
convincingly by the convergence of the results as function of $m$. 
A ln-ln plot of the raw data for the gap as a function of chain length $N$ 
is not very revealing, but rather misleading as has been pointed out by Carlon et al. 
\cite{Carlon}. A much more refined way of analyzing the data is to use effective exponents. 
For the gap we define  
\begin{equation} \label{d1}
z_N = {\ln \tau(N+1) - \ln \tau(N-1) \over \ln (N+1) - \ln (N-1)} 
\simeq {d \ln \tau \over d \ln N}.
\end{equation}
which is a function of the chain length. If the renewal time were a strict power law 
$\tau \sim N^z$, the expression (\ref{d1}) would equal $z$,  independent of $N$. We want to 
stress that only very smooth data can be used to calculate these effective exponents, 
because noise, which is e.g. inherent in simulations, will magnify in the ratio of small 
differences. 

In order to get an estimate of the convergence we give in Fig.~\ref{figCG06} two sets 
of curves for $c=0.01$ and $c=0$ for various $m$. The curves for $c=0.01$ are quite close,
such that further increase of $m$, does not lead to significantly different results. 
On the contrary, for the pure reptation case $c=0$, the curves keep changing  with $m$ 
for longer and longer chains. 
Note that the basis is already unusually large ($m=180$) for DMRG calculations. The 
large $m$ was possible due to the 
speed-up of the process by using the full symmetry of the lattice. The double symmetry
allowed to enlarge the size of the basis $m$ by a factor 4.
For each value of $c$ there is a maximal value of $N$ for which the result converge
for feasible values of $m$. Within that range the 
DMRG values are also sufficiently accurate such that the small differences in  
(\ref{d1}) do not suffer from computational noise. In the pictures of the coming sections 
we only plot the data which do not depend on the value of $m$.

For the diffusion coefficient the domain of $m$-independent data is even more restricted.
Here we introduce, similar to (\ref{d1}), the effective diffusion exponent
\begin{equation} \label{d2}
x_N = -{\ln D(N+1) - \ln D(N-1) \over \ln (N+1) - \ln (N-1)} 
\simeq -{d \ln D \over d \ln N}.
\end{equation} 
In Fig.~\ref{figCG07} we show the exponent $x_N$ for $c=0$ and $c=0.01$. The latter is 
again reasonably convergent, but for the former we could not go to sufficiently large 
$m$ such that a convergent domain starts to emerge. One also observes noise 
which is not visible on a ln-ln plot, but which shows up as a result of small numbers in 
numerator and denominator in (\ref{d2}). So practically our calculations are limited to 
values larger than $c=0.001$. 

\begin{figure}[h]
\begin{center}
    \epsfxsize=12cm
    \epsffile{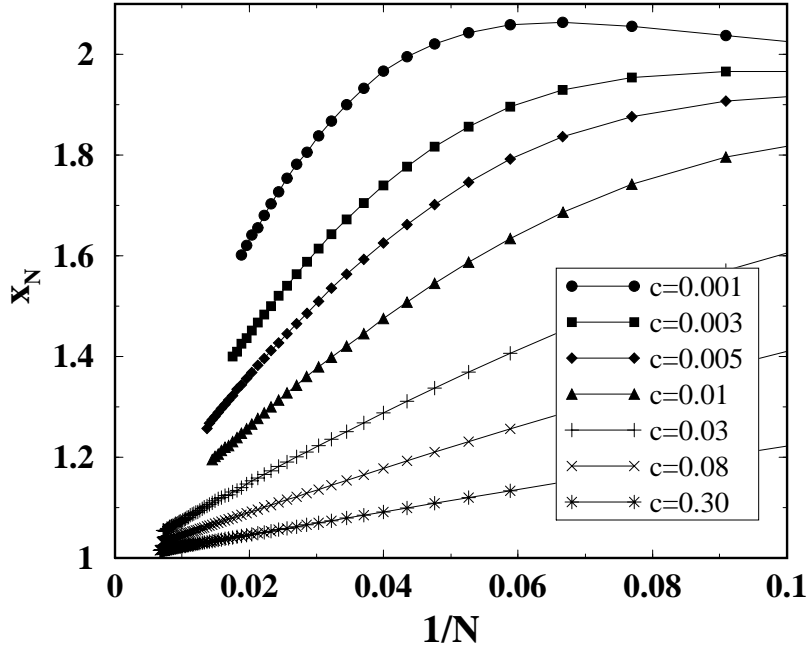}
    \caption{The diffusion exponent as a function of the length of the chain for various
    values of the barrier crossing rate $c$.}  \label{figCG02}
\end{center}
\end{figure}

The local exponents $z_N$ for the renewal time obtained for various $c$ are collected in Fig.~\ref{figCG01}.
In Fig.~\ref{figCG02} we plot, in the 
same way, the local exponent $x_N$ for the diffusion coefficient. Generally, in spite of the above
mentioned restrictions, we can make the following observations.
\begin{itemize}
\item Chains of the order of $N \simeq 100$ are not yet in the asymptotic regime. The
effective exponents still deviate appreciably from the asymptotic value. In other words,
there are large corrections to scaling.  
In particular, the plateaus in the small $c$ curves may easily lead to the conclusion 
that the exponent has settled on a too large value. 
\item The influence of small values of $c$ is quite strong for long chains.
We come back on this point when we discuss the crossover behavior.
\item Although we have no clear evidence that the $c=0$ curve for the gap tends towards
the asymptotic value $z_\infty = 3$, it is clear that the curves for smaller and smaller
$c$ ``try'' to approach this theoretical value for reptation. The approach
to the asymptotic value $z_\infty = 2$, for larger $c$, is evident. This is the
exponent for Rouse dynamics. 
\item The curves that we could calculate for the diffusion coefficient
approach the asymptotic exponent $x_\infty =1$, which is again the Rouse exponent for
diffusion. 
\end{itemize}

\begin{figure}[h]
\begin{center}
    \epsfxsize=12cm
    \epsffile{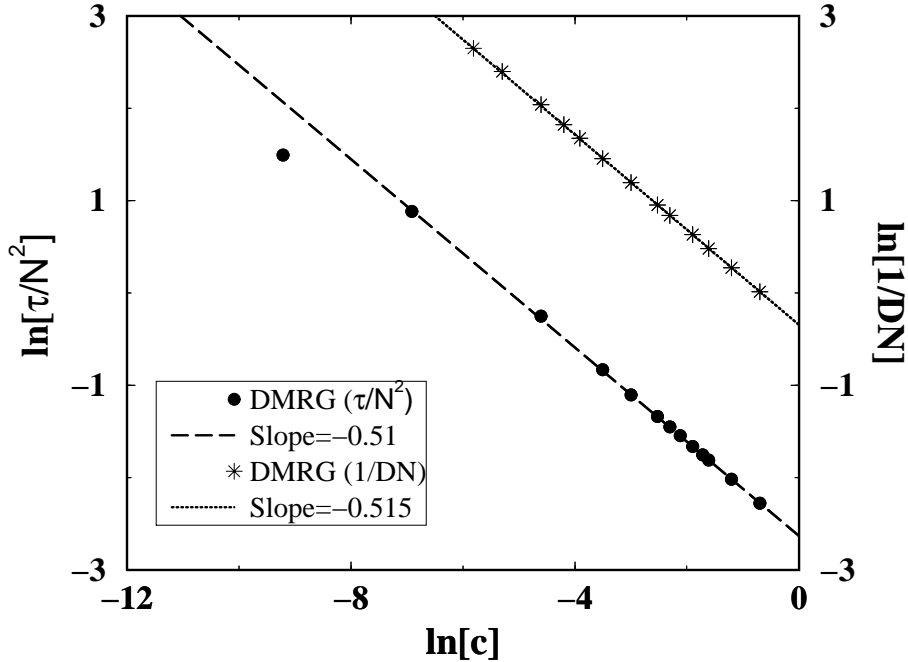}
    \caption{Ln-ln plots of the renewal time and the diffusion coefficient as 
    function of c.}  \label{figCG03}
\end{center}
\end{figure}

\section{Crossover scaling}

As the curves of Figures \ref{figCG01} and \ref{figCG02} show, the renewal time
$\tau$ and the diffusion coefficient $D$ are widely varying functions of the two 
parameters $c$ and $N$.
We can organize the data more transparently in terms of a crossover
scaling function, aiming at data collapse.
\begin{figure}[h]
\begin{center}
    \epsfxsize=12cm
    \epsffile{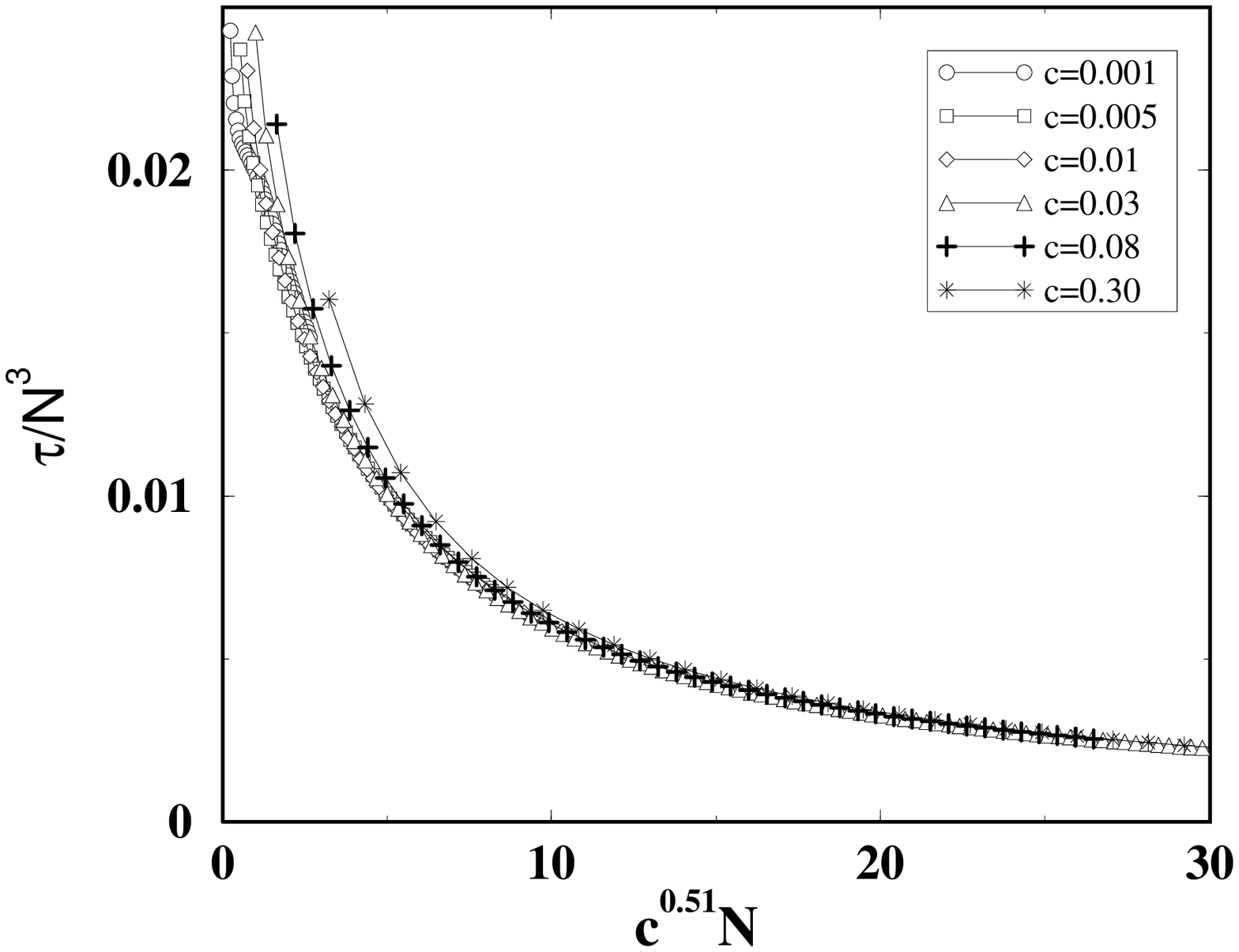}
    \caption{The crossover function $g$.
    }  \label{figCG04}
\end{center}
\end{figure}
Anticipating the asymptotic values of the two regimes: 
$c \rightarrow 0$ and a fixed $c \neq 0$, the following representation is adequate
for the renewal time
\begin{equation} \label{e1}
\tau (N, c) = N^3 g (c^\theta N).
\end{equation} 
From such a representation one derives for the effective exponent the expression
\begin{equation} \label{e2}
{d \ln \tau \over d \ln N} = 3 + {d \ln g (c^\theta N) \over d \ln (c^\theta N)}.
\end{equation} 
The crossover function $g(x)$ itself should be expandable for small arguments as
\begin{equation} \label{e3}
g(x) = g_0 + g_1 x + \cdots
\end{equation} 
and for large arguments as
\begin{equation} \label{e4}
g(x) \simeq {1 \over x} \left( g_{-1} + {g_{-2} \over x} + \cdots \right).
\end{equation}
Inserting the asymptotic behavior (\ref{e4}) into (\ref{e1}) we obtain
\begin{equation} \label{e5}
\ln(\tau/N^2) = \ln g_{-1} - \theta \ln c + \cdots,
\end{equation} 
where the dots refer to corrections of order $1/N$. In Fig.~\ref{figCG03} we have made a 
plot of $\ln(\tau/N^2)$ vs $\ln c$, extrapolated to $N \rightarrow \infty$, which 
corresponds to the first two terms of (\ref{e5}).
As one sees the curve is fairly straight, with a slope $ \simeq -0.51$, 
in the domain where the data are most accurate. 
We use this value in a scaling plot of $g(x)$,
which is shown in Fig.~\ref{figCG04}. The most important part of the figure is that for
large argument the data collapse and fall on a curve which decays as $1/x$, implying
the crossover from reptation to Rouse dynamics. The pure reptation behavior
follows from the finiteness of $g(x)$ for small argument.
The value $g(0)$ can be derived from a plot of $\tau N^{-3}$ versus $N^{-1}$. 
We find the value $g(0) \simeq  0.026$, which is in good agreement with the behavior
of the curves for small argument. We see that the data at small $x$ do not coincide as
good as for large $x$. The reason is the inclusion of small values of $N$ for small $c$.
Here the $N$ is not sufficiently large to have scaling. One should make $N$ larger and 
$c$ smaller to get good scaling in that region, but that regime is as yet inaccesible to us. 

\begin{figure}[h]
\begin{center}
    \epsfxsize=12cm
    \epsffile{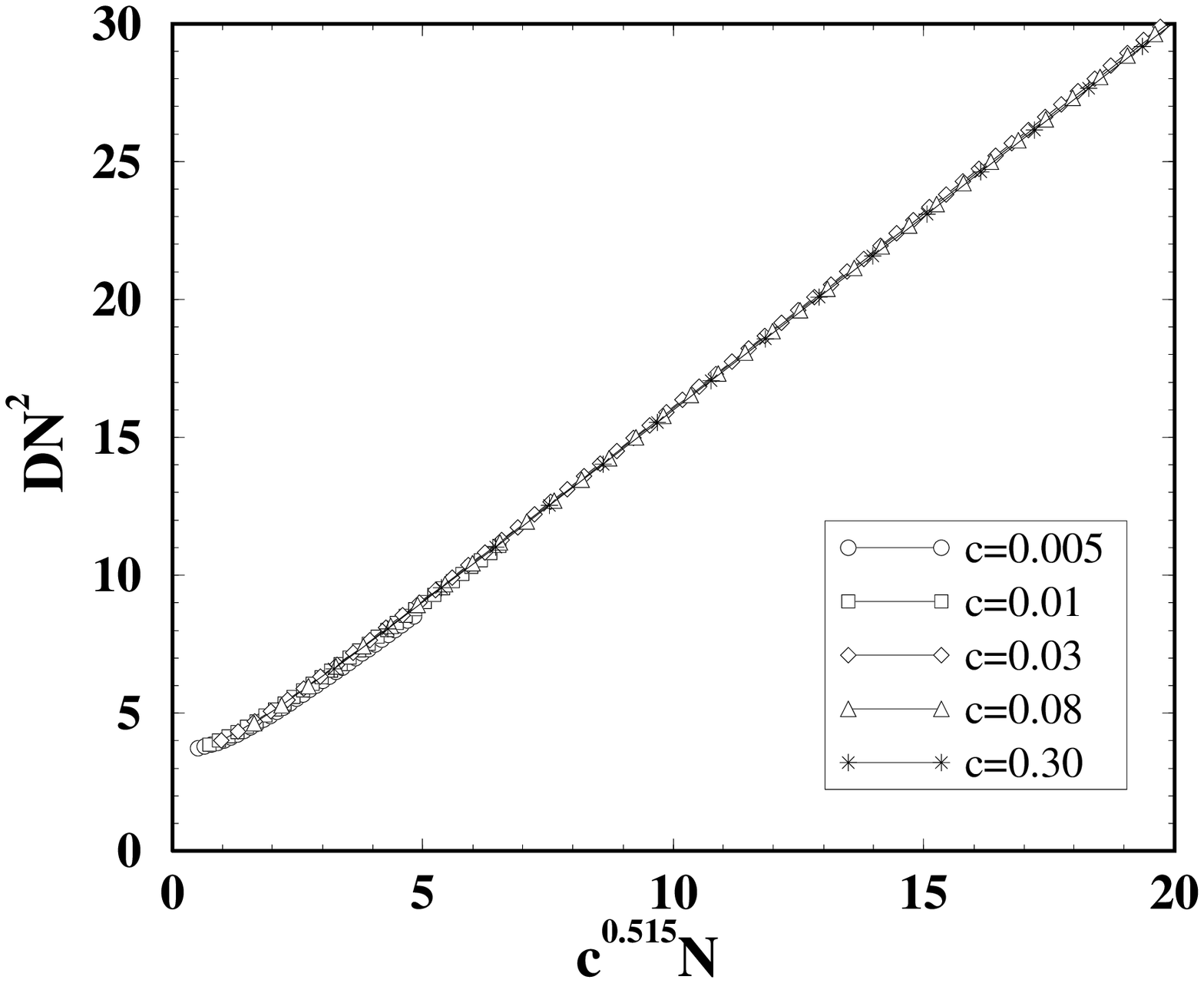}
    \caption{The crossover function $f$.
    }  \label{figCG05}
\end{center}
\end{figure}
In Fig.~\ref{figCG03} also $\ln 1/D N$ is plotted as function of $\ln c$. It gives again
a straight line with the same slope $\theta \simeq 0.515$. We use this value in the scaling
plot, Fig.~\ref{figCG05}, for the diffusion coefficient in the form
\begin{equation} \label{e6}
D (N, c) = N^{-2} f(c^\theta N).
\end{equation}
As one sees the collapse is here amazingly
good in view of the lesser quality of the data, as compared to the renewal time.
Also smaller $N$ and ``large'' $c$ are included, although
no real scaling can be expected for these values. It proves that 
crossover scaling works very well for the diffusion coefficient.
The crossover scaling function $f$ approaches again a finite value at $x=0$ that
can be estimated from Fig.~\ref{figCG05} as $f(0) \simeq 3.67$. 
For large arguments, $f(x)$ should behave as $f(x \rightarrow \infty) \sim x$, which is
confirmed by the plot. It shows again the crossover from reptation to Rouse dynamics.

\section{Discussion}

We have found that the renewal time and the diffusion coefficient can be transparently 
described by the crossover scaling functions (\ref{e1}) and (\ref{e6}). In particular the 
diffusion data fit the scaling curve very well for practical all the points calculated. 
The crossover scaling exponent is found to be $\theta = 0.51$ and we are fairly convinced 
that the exact value is $1/2$. Not only the data support this value but also an analytical 
argument can be given for $\theta = 1/2$, which runs as follows. 
Reptation does not change the backbone of the chain (which results from stripping the 
hernias from the chain). The hernias walk along the backbone and are created and 
annihilated at the end of the chain. For the removal of a backbone segment of 
length $N$, the end repton has to diffuse over a distance of order $N$. For this 
diffusion along the backbone the curvilinear diffusion coefficient applies, which is an 
order $N$ larger than the total diffusion coefficient, so it is of order $N^{-1}$ 
(see also section \ref{deq1}). The time scale for diffusion is distance squared ($N^2$) 
divided by the diffusion constant ($N^{-1}$). Thus a backbone segment of order $N$ 
requires a time scale $N^2 /N^{-1}  = N^3$ to be renewed (this in fact explains the 
reptation exponent $z=3$). On the other hand, to change that backbone segment by 
direct hopping over barriers, one needs the time $N/c$. The fastest process dominates 
and the competition is controlled by the ratio of the rates $N^3/(N/c) = cN^2$. So the 
crossover scaling function should be a function of the ratio $c N^2$, 
which yields $\theta=1/2$. 

One may wonder why it is so difficult for DMRG to reach long $N$ for small $c$, while the
hamiltonian simplifies for $c=0$. As one observes from the figures \ref{figCG01} and 
\ref{figCG02}, there is a zone where the values make a turn from reptative behavior to
Rouse dynamics. The basis in the DMRG approximation has to be large enough to notice this
difference in behavior at the appropriate $N$, which grows as $c^{-1/2}$. One needs
a basis size $m$ of the same order to describe the crossover. It does not mean that it is
impossible to use DMRG for the pure cage model with $c=0$. But then one has to use a 
representation of the hamiltonian which explicitly acknowledges the extra symmetries which
are present in this model. Additional symmetries of the cage model are discussed in the 
papers of van Heukelum et al. \cite{Heukelum}. 
A similar situation occurs in crossover in the Rubinstein-Duke model \cite{drzewin2}. 

The results of this paper are quite similar to the crossover found in a one-dimensional 
Rubinstein-Duke model with hernia creation and annihilation \cite{Drzewinski}. 
This leads us to believe that the crossover in gels is a universal phenomenon with the
crossover exponent $\theta =1/2$, independent of the embedding lattice, as long as the
embedding lattice permits sideways motion, which cross the barriers. Sometimes the motion 
rules exclude crossing of barriers, e.g. in the one-dimensional embedding. But also
in a lattice with triangular cells, crossing is impossible within the rule that links
are always between nearest neighbor cells.

We find that with fixed non-zero crossing rate $c$, the chain always tends towards Rouse
dynamics for larger and larger $N$. This contrasts the general observation that in polymer
melts the opposite tendency takes place: longer chains display reptative behavior 
\cite{Kremer2, Wischnewski}.
It is clear that the obstruction due to other polymers cannot be seen as a fixed barrier,
with a certain tunneling rate. Thus our results cannot be applied 
to polymer melts using a fixed rate for sideways motion.
In other words, $c$ must become a function of the chain length. To handle a chain length 
dependent hamiltonian gives a complication in DMRG. Apart from the difficulties to find  
an adequate model for polymer melts that allows to treat very long chains accurately, we 
may speculate that an ``effective'' rate $c$ for sideways motion depends as a power 
$N^{-\alpha}$ on the length $N$. The combination $c N^2 \sim N^{2-\alpha}$ determines 
whether one sees reptation of Rouse dynamics. It is tempting to take for $\alpha$ the 
effective renewal exponent $z_N$, defined in (\ref{d1}), because it takes the renewal time 
for a polymer to get out of the way. As this $\alpha$ is always larger than 2, the combination 
shrinks with growing length, making the $\alpha$ even larger.  Thus one observes the 
opposite crossover: from Rouse dynamics to reptation.    

{\bf Acknowledgement} The authors like to thank Gerard Barkema for most stimulating 
discussions.


\begin{thebibliography}{99}
\bibitem{footnote} One could rename $B^2$ to $B$, but this might give
confusion with respect to the convention in the Rubenstein-Duke model, where moves with
bias $B$ are allowed.
\end{thebibliography}
\end{document}